\documentclass{biophys_letter}
\usepackage{dcolumn,color,helvet,times}
\usepackage{amsmath,lastpage,bm,textcomp}
\usepackage{multicol}
\jno{kxl014} 
\gridframe{N}
\cropmark{N}
        
\doi{doi: 10.1016/bpj.xxxxxxxxx}

\begin{document}
\renewcommand{\thefootnote}{\fnsymbol{footnote}}

\setcounter{page}{1} 
\title{Chain Length Determines the Folding Rates of RNA}

\author{Changbong Hyeon,*and D. Thirumalai{\authdagger}}

\address{*School of Computational Sciences, Korea Institute for Advanced Study, Seoul 130-722, Korea and {\addrdagger}Biophysics Program, Institute for Physical Science and Technology,
University of Maryland, College Park, MD 20742}

\maketitle

\pagestyle{headings}

\markboth{Biophysical Journal: Biophysical Letters}{Biophysical Journal: Biophysical Letters} 


\begin{abstract}
{We show that the folding rates ($k_F$s) of RNA are determined by $N$, the number of nucleotides.  By assuming that the distribution of free energy barriers separating the folded and the unfolded states is Gaussian, which follows from central limit theorem arguments and polymer physics concepts, we show that $k_F \approx k_0 \exp{(-\alpha N^{0.5})}$. Remarkably, the theory fits the experimental rates spanning over seven orders of magnitude with $k_0 \sim 1.0 (\mu s)^{-1}$. An immediate  consequence of our finding is that the speed limit of RNA folding is about one microsecond just as it is in the folding of globular proteins.  }
{Received for publication Dec 2011 and in final form x xxxx xxxx.}{Address reprint requests and inquiries to C. Hyeon or D. Thirumalai, E-mail: hyeoncb@kias.re.kr, thirum@umd.edu}
\end{abstract}

\vspace*{2.7pt}
\begin{multicols}{2}

RNA molecules are evolved biopolymers whose folding has attracted a great deal of attention \cite{Thirum05Biochem,TreiberCOSB99,Woodson10ARB} because of the crucial role they play in a number of cellular functions. The slightly branched polymeric nature of RNA implies that the shapes, relaxation dynamics, and even their folding rates must depend on $N$.  In support of this assertion it has been shown that the radius of gyration of the folded states, using data available in Protein Data Bank (PDB), scales $R_g \sim 5.5 N^{\nu}$ \AA, where the Flory exponent $\nu$ varies from 0.33 to 0.40 \cite{Hyeon06JCP_2,Yoffe08PNAS,Hajdin10RNA}.   
Although 
this result is expected from the  perspective of polymer physics 
it is surprising from the viewpoint of structural biology because it might be argued that sequence and the complexity  of secondary and tertiary structure organization could lead to substantial deviations from the predictions based on Flory-like theory. Here, we show that folding rates, $k_F$s,  of RNA are also primarily determined by $N$, thus adding to the growing evidence that it is possible to understand  folding of RNA using polymer physics principles.
\\

\noindent \textbf{Theoretical Considerations.}
Theoretical arguments, with genesis in the dynamics of activated transitions in disordered systems, suggest that 
\begin{equation}
k_F = k_0 \exp{(-\alpha N^{\beta})}
\end{equation} 
where $\beta$ should be 0.5 \cite{Thirum95JPI}. 
The rationale for this finding hinges on the observation that favorable base-pairing interactions and the  hydrophobic nature of the bases tend to collapse RNA whereas the charged phosphate residues are better accommodated by extended structures. Thus, the distribution of activation free energy, $\Delta G^{\ddagger}_{UF}/k_BT$,  between the folded and unfolded states is a sum of favorable and unfavorable terms. We expect from central limit theorem that the distribution of $\Delta G^{\ddagger}_{UF}/k_BT$ should be roughly Gaussian with dispersion $\langle (\Delta G^{\ddagger}_{UF})^2\rangle\sim N$.   Thus,  $\Delta G^{\ddagger}_{UF}/k_BT\sim N^{\beta}$  with $\beta=1/2$. 

We analyzed the available experimental data (see the Table for a list of RNA molecules) on RNA folding rates by assuming  that $\Delta G_{UF}^{\ddagger}$ grows as $N^{\beta}$ with $\beta$ as a free parameter.  The theoretical value for $\beta$ is 0.5.   The folding rates of RNA spanning over \textit{seven} orders of magnitude is well fit using $\log k_F=\log k_0-\alpha N^{\beta}$ with correlation coefficient of 0.98 (Fig.1). The fit  yields $k_0^{-1}=0.87$ $\mu s$, $\alpha=0.91$ and $\beta \approx 0.46$.  In the inset we show the fit obtained by fixing $\beta = 0.5$. Apart from the moderate differences in the $k_0^{-1}$ values the theoretical prediction and the numerical fits are in agreement, which demonstrates that the major determining factor in determining RNA folding rates is $N$. 

It is known that RNA, such as \emph{Tetrahymena} ribozyme, folds by multiple pathways that is succinctly described by the kinetic partitioning mechanism (KPM) \cite{Guo95BP}.  
According to KPM a fraction, $\Phi$, of molecules reaches the native states rapidly whereas the remaining fraction is trapped in an ensemble of misfolded intermediates. For \emph{Tetrahymena} ribozyme $\Phi \sim 0.1$ \cite{PanJMB97}.  The $N$-dependence given by Eq. (1) holds for the majority of molecules that fold to the native state from the compact intermediates, which form rapidly under folding conditions \cite{ThirumARPC01}.  
\\

\noindent \textbf{Conclusions.}
Implications of our findings are: (i) The inverse of the prefactor, $k_0^{-1}=\tau_0\approx 0.87$ $\mu s$, is almost six orders of magnitude larger than the transition state theory estimate of $h/k_BT\approx 0.16$ $ps$. 
The value of $\tau_0$, which  coincides with the typical base pairing time \cite{porschke1973BP}, is the speed limit for RNA folding.  Interestingly, general arguments based on the kinetics of loop formation have been used to predict that the speed limit for protein folding is also about one $\mu s$ \cite{Hagen96PNAS,Yang03Nature,Eaton04COSB}. It remains to be ascertained 
if the common folding speed limit for proteins and RNA is due to evolutionary pressure on the folding of evolved sequences. 
\doiline
\end{multicols}
\twocolumn
\begin{figure}[t!]\vspace*{-5pt}
\centering{\includegraphics[width=3.30in]{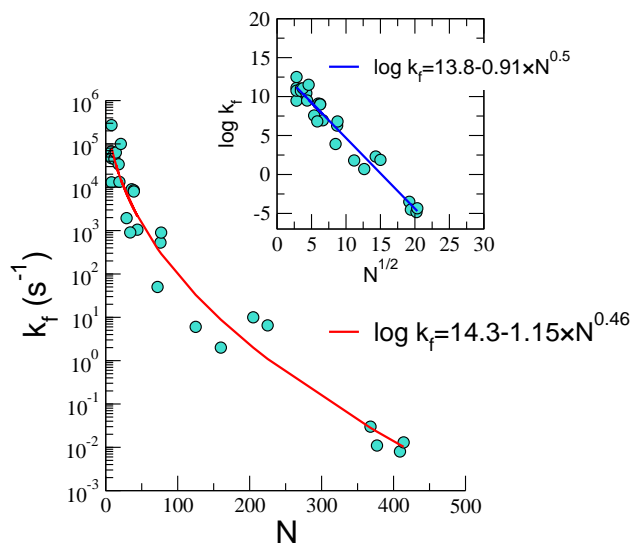}}\vspace*{-3pt}
\caption{Dependence of folding rates of RNA on $N$. The green circles are experimental data and the red line is the fit using $\log{k_F}=\log{k_0}-\alpha N^{\beta}$ using $\beta$ as an adjustable parameter. The inset shows the fit obtained by fixing $\beta$ to the predicted theoretical value of 0.5. }
\vspace*{-1.5pt}
\end{figure}  
\noindent (ii) It is worth pointing out that Dill \cite{Dill11PNAS} has recently shown that the dependence of  rates and stabilities of protein folding depend only the number of amino acids, which in turn places strict constraints on their functions in the cellular context. Taken together these studies show that despite the complexity of protein and RNA folding only a few variables might determine their global properties, which suggests that there may be simple principles that determine biological functions.  
\\

\noindent \textbf{Acknowledgements.} This was supported by a grant from the National Science Foundation through grant number CHE 09-14033.
\\


\onecolumn 
\begin{table}
\begin{minipage}{\textwidth}
\centering 
\caption[Caption for LOF]%
     {RNA length versus Folding rate.
     }
  {\footnotesize
  \begin{tabular}{|c||c|c|c|}    \hline                                   
  RNA &N
  &$k_f(sec^{-1})$
  &\\
    \hline\hline
    GCUUCGGC \cite{Proctor04BC}                         &               8 &    $6.7\times 10^4$& tetraloop hairpin\\
    GCUUCGGC \cite{Proctor04BC}                         &               8 &    $27.2\times 10^4$&tetraloop hairpin \\
    GGUUCGCC \cite{Proctor04BC}                         &               8 &    $1.3\times 10^4$&  tetraloop hairpin\\
    GGUUCGCC \cite{Proctor04BC}                         &               8 &    $4.7\times 10^4$&  tetraloop hairpin\\
    GGACUUUUGUCC \cite{Proctor04BC}                     &              12 &    $6.1\times 10^4$& tetraloop hairpin \\
    GGACUUCGGUCC \cite{Proctor04BC}                     &              12 &    $4.5\times 10^4$& tetraloop hairpin \\
    A$_6$C$_6$U$_6$ \cite{Porschke74BPC}                &              18 &    $3.4\times 10^4$&  tetraloop hairpin\\
    Extra-arm of $\mathrm{tRNA^{Ser}}$(yeast) \cite{Riesner}&               21&      $1\times 10^5$& tRNA \\
    pG-half of $\mathrm{tRNA^{Phe}}$(yeast) \cite{Riesner}&               36&      $9\times 10^3$&  tRNA \\
    CCA-half of $\mathrm{tRNA^{Phe}}$(yeast) \cite{Riesner}&               39&    $8.5\times 10^3$& tRNA\\
    CCA-half of $\mathrm{tRNA^{Phe}}$(wheat) \cite{Riesner}&               39&      $8\times 10^3$& tRNA\\
    $\mathrm{tRNA^{Phe}}$(yeast) \cite{serebrov2001Biochem}&               76&      $5.3\times 10^2$    &          tRNA\\
    $\mathrm{tRNA^{Ala}}$(yeast) \cite{Riesner}&               77&      $9\times 10^2$     &     tRNA    \\
    $Y_4$-hairpin \cite{kuznetsov2008NAR} & 14 &$5.75\times 10^4$&hairpin (5$\times$2+4) \\
    $Y_9$-hairpin \cite{kuznetsov2008NAR} & 19 &$2.29\times 10^4$& hairpin (5$\times$2+9)\\
    $Y_{19}$-hairpin \cite{kuznetsov2008NAR} & 29 &$8.70\times 10^2$& hairpin (5$\times$2+19)\\
    $Y_{34}$-hairpin \cite{kuznetsov2008NAR} & 44 &$6.03\times 10^2$& hairpin (5$\times$2+34)\\
    VPK pseudoknot \cite{Narayanan11JACS} & 34& $9.09\times 10^2$& pseudoknot\\
     Hairpin ribozyme (4-way junction) \cite{Tan03PNAS,Sosnick04JMB}& 125 & $6$& natural form of hairpin ribozyme \\
    P5abc \cite{DerasBIOCHEM2000}                         &               72&   $50$& Group I intron $T$. ribozyme\\
    P4-P6 domain(\emph{Tetrahymena} ribozyme) \cite{DerasBIOCHEM2000}&            160&    $2$ & Group I intron $T$. ribozyme\\
    Azoarcus \cite{RanganPNAS2003,Sosnick04JMB}                      &              205&    $7\sim 14$& \\
    B.subtilis RNase P RNA catalytic domain \cite{FangNSB99}      &              225&         $6.5\pm0.2$& \\
    Ca.L-11 ribozyme \cite{ZhangNAR03}                           &              368&   $0.03$         &   \\
    E.coli RNase P RNA \cite{WilliamsonRNA96}           &              377&   $0.011\pm0.001$ & \\
    B.subtilis RNase P RNA \cite{WilliamsonRNA96}&              409&    $0.008\pm0.002$  &\\
    \emph{Tetrahymena} ribozyme \cite{SclaviSCI98, Sosnick04JMB}      &              414&            $0.013$  &Group I intron $T$. ribozyme \\
    \hline
  \end{tabular}}
\label{table:RNA_rate}
\end{minipage}
\end{table}


\end{document}